\documentclass[lettersize,journal]{IEEEtran}

\usepackage{graphicx}
\usepackage{booktabs}
\usepackage{siunitx} 
\usepackage{amsmath,amssymb,amsfonts}
\usepackage{array}
\usepackage{tabularx}
\usepackage{enumitem}
\usepackage{makecell}
\usepackage{listings}
\usepackage{xcolor}
\usepackage{tikz}
\usepackage{pgfplots}
\pgfplotsset{compat=1.18}

\usepackage{balance}

\usepackage[colorlinks=true,linkcolor=black,citecolor=black,urlcolor=black]{hyperref}

\definecolor{codegreen}{rgb}{0,0.6,0}
\definecolor{codegray}{rgb}{0.5,0.5,0.5}
\definecolor{codepurple}{rgb}{0.58,0,0.82}
\definecolor{backcolour}{rgb}{0.95,0.95,0.92}
\definecolor{codered}{rgb}{0.91,0.58,0.53}

\definecolor{softblue}{RGB}{102,178,255}
\definecolor{softred}{RGB}{255,153,153}
\definecolor{softgreen}{RGB}{153,255,153}
\definecolor{softyellow}{RGB}{255,255,153}
\definecolor{softcyan}{RGB}{153,255,255}
\definecolor{softmagenta}{RGB}{255,153,255}
\definecolor{softorange}{RGB}{255,200,150}

\def\BibTeX{{\rm B\kern-.05em{\sc i\kern-.025em b}\kern-.08em
    T\kern-.1667em\lower.7ex\hbox{E}\kern-.125emX}}

\lstdefinestyle{mystyle}{
    backgroundcolor=\color{backcolour},   
    commentstyle=\color{black},
    keywordstyle=\color{black},
    numberstyle=\scriptsize\color{codegray},
    stringstyle=\color{black},
    basicstyle=\ttfamily\scriptsize, % Set the basic style to scriptsize font size
    breakatwhitespace=false,         
    breaklines=true,                 
    captionpos=b,                    
    keepspaces=true,                 
    numbers=left,                    
    numbersep=5pt,                  
    showspaces=false,                
    showstringspaces=false,
    showtabs=false,                  
    tabsize=2
}

\lstdefinestyle{mystyle_1}{
    backgroundcolor=\color{codered},   
    commentstyle=\color{codegreen},
    keywordstyle=\color{black},
    numberstyle=\scriptsize\color{codered},
    stringstyle=\color{codered},
    basicstyle=\ttfamily\scriptsize, % Set the basic style to scriptsize font size
    breakatwhitespace=false,         
    breaklines=true,                 
    captionpos=b,                    
    keepspaces=true,                 
    numbers=left,                    
    numbersep=5pt,                  
    showspaces=false,                
    showstringspaces=false,
    showtabs=false,                  
    tabsize=2
}

\newcommand{\tool}{\textsc{Xamt}}

\begin{document}

\title{Testing Deep Learning Library APIs via Cross-Framework Differential Fuzzing}

\author{
Bin Duan, 
Ruican Dong, 
Naipeng Dong, 
Dan Dongseong Kim, 
Guowei Yang* \\
School of Electrical Engineering and Computer Science, \\
The University of Queensland, Brisbane, Australia \\
b.duan@uq.edu.au, ruican.dong@uq.net.au, n.dong@uq.edu.au, dan.kim@uq.edu.au, guowei.yang@uq.edu.au
}

\maketitle

\begingroup
\renewcommand\thefootnote{*}
\renewcommand{\footnoterule}{}
\footnotetext{Corresponding author.}
\endgroup

\begin{abstract}
Deep learning libraries underpin many safety- and reliability-critical applications, yet existing API-level testing techniques often rely on intra-library properties or CPU--GPU differential oracles and may miss defects that behave consistently across hardware backends. We present \tool, a cross-framework differential fuzzing approach for deep learning library APIs. \tool constructs and tests execution-validated groups of APIs intended to implement equivalent operations across seven libraries. It uses explicit API aliases and parameter-role normalization to construct candidate correspondences and validates them through pairwise execution and a group-level behavioral check on canonical ordinary inputs. The resulting groups are explored using variance-guided differential fuzzing with ordinary, boundary, and non-finite inputs. Crash and inconsistency oracles flag executions exhibiting abnormal termination or inconsistent outputs for subsequent reproduction and analysis. Across the seven libraries, \tool\ constructs 676 execution-validated groups containing 2,563 matched APIs. Among these, \tool\ identifies 72 independently reproduced discrepancy cases, including 4 crash cases and 68 output inconsistencies. Among the 72 developer reports, 25 have been confirmed, including 23 that have been fixed.
\end{abstract}


\section{Introduction}

Deep learning has become integral to real-world systems in domains such as autonomous driving~\cite{rao2018deep,bogdoll2022anomaly}, healthcare~\cite{miotto2018deep,arabahmadi2022deep}, and finance~\cite{heaton2017deep,venkateswarlu2022efficient}. These applications depend heavily on deep learning libraries, whose APIs provide the building blocks for model construction, training, and inference. Implementation defects in these APIs, including incorrect numerical results, inconsistent handling of non-finite values, and silent logic errors, can propagate to downstream models and compromise the reliability of deployed systems.

Recent studies apply fuzzing to detect bugs in deep learning libraries by generating diverse inputs and using test oracles to identify abnormal behavior~\cite{li2018fuzzing,xie2022docter}. Existing techniques can be broadly divided into model-level and API-level testing. Model-level fuzzing~\cite{guo2020audee,gu2022muffin,wang2020deep} generates or mutates complete neural networks and observes their behavior under different execution settings. API-level fuzzing~\cite{wei2022free,deng2023largetitan} directly explores the input domains of individual library APIs. Both directions have revealed defects in widely used deep learning infrastructure.

Despite these advances, many existing approaches rely on intra-library properties or differential oracles, particularly comparisons between CPU and GPU executions. Such oracles are effective when a defect manifests differently across hardware backends, but they provide no disagreement signal when an implementation produces the same divergent result on both CPU and GPU. Consequently, an API may behave consistently across backends while still returning an incorrect result. Cross-framework testing provides a complementary perspective by comparing independently implemented APIs intended to perform the same operation.

To address this problem, we present \tool, a cross-framework differential fuzzing approach for deep learning library APIs. \tool\ constructs candidate correspondences using explicit API aliases and parameter-role normalization, aligns library-specific interfaces through shared logical inputs, and retains supported correspondences after pairwise and group-level execution validation. It then applies variance-guided fuzzing over ordinary, boundary, and non-finite inputs.

We evaluate \tool\ on PyTorch, TensorFlow, Keras, JAX, MindSpore, PaddlePaddle, and Chainer. Across these seven libraries, \tool\ constructs 676 execution-validated cross-library groups spanning 74 distinct library combinations and containing 2,563  matched APIs. The testing campaign identifies 72 independently reproduced discrepancy cases, including four Crash cases and 68 output inconsistencies. Among the 72 developer reports, 25 have been confirmed, including 23 that have already been fixed. Moreover, among the 50 CPU--GPU-applicable cases in the backend-replay subset, none produces an observable backend disagreement. These results show that the same triggering inputs provide no CPU--GPU disagreement signal under the evaluated configurations.

This article is an extended version of our earlier paper presented at IEEE ISSRE~\cite{xamt}. The extension has two parts. Methodologically, \tool\ removes the designated-reference-library requirement through API aliases and parameter-role normalization, adds pairwise and group-level execution validation of candidate correspondences, and broadens input generation to include boundary and non-finite values. Experimentally, we expand the evaluation from five to seven libraries and add analyses of matching quality, fixed-input CPU--GPU replay, method-level rediscovery by existing baselines, TensorScope correspondence coverage, component contributions, and testing-budget sensitivity.

\begin{figure}[t!]
  \centering
  \includegraphics[width=1\linewidth]{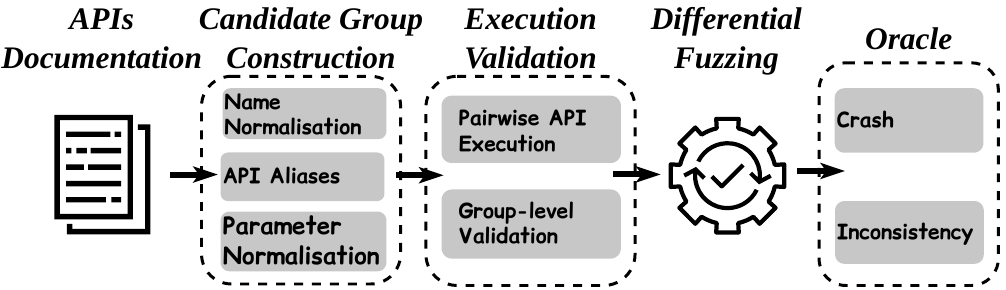}
  \caption{Overview of \tool.}
  \label{fig:overview}
\end{figure}

\section{Approach}
\label{sec:approach}

Fig.~\ref{fig:overview} presents an overview of \tool. First, \tool\ collects callable APIs from the configured namespaces of the seven target libraries and normalizes their operation names and parameter roles. APIs with the same normalized operation name are placed into candidate groups, while coarse namespace-derived categories are used only to exclude obvious semantic mismatches. Second, \tool\ aligns their input parameters and validates each supported cross-library API pair through actual execution on canonical-operation-specific ordinary inputs. Pairs that execute successfully and produce consistent outputs are connected to form candidate components, which are further validated at the group level. Third, \tool\ generates consistent inputs and applies variance-guided differential fuzzing within a fixed time budget. Finally, Crash and Inconsistency oracles flag executions exhibiting abnormal termination or inconsistent outputs for subsequent reproduction and analysis.

\subsection{Defining API Matching Rules}
\label{sec:matching-rules}

We introduce explicit API aliases and parameter-role normalization to accommodate library-specific naming and interface differences and to reduce ambiguous matches. These rules are constructed by comparing API names, signatures, and parameter descriptions in the official documentation of the seven target libraries and consolidating recurring library-specific variants.

\subsubsection{API Name Normalization and Aliases}

API names often reflect their core functionality, but different libraries may use substantially different names for the same operation. \tool\ first converts API names from camel case to lowercase underscore-separated identifiers, normalizes special characters and dimension suffixes, and removes configured library and namespace prefixes. Version- and backend-specific suffixes, such as \texttt{\_v2} and \texttt{\_eager\_fallback}, are also removed when applicable.

After syntactic normalization, \tool\ applies a curated alias mapping that converts library-specific names into common operation names. We construct the mapping by comparing API names, callable signatures, parameter descriptions, and operation semantics in the official documentation of the seven target libraries. An alias is retained only when the corresponding APIs implement the same canonical operation within an identifiable shared semantic domain. Table~\ref{tab:api-aliases} presents representative examples. For example, \texttt{torch.clamp} and \texttt{tensorflow.clip\_by\_value} are both normalized to \texttt{clip}, while \texttt{torch.unsqueeze} and \texttt{jax.numpy.expand\_dims} are normalized to \texttt{expand\_dims}. These mappings allow corresponding APIs to enter the same candidate group even when their original names have limited lexical similarity.

\begin{table}[t]
\centering
\footnotesize
\caption{Representative API Name Aliases Used by \tool.}
\label{tab:api-aliases}
\setlength{\tabcolsep}{4pt}
\renewcommand{\arraystretch}{0.85}
\begin{tabularx}{\columnwidth}{@{}l X@{}}
\toprule
\textbf{Normalized Name} & \textbf{Library-Specific Names} \\
\midrule
\texttt{concatenate} 
& \texttt{cat}, \texttt{concat} \\
\texttt{clip} 
& \texttt{clamp}, \texttt{clip\_by\_value} \\
\texttt{arccos} 
& \texttt{acos} \\
\texttt{sum} 
& \texttt{reduce\_sum} \\
\texttt{matmul} 
& \texttt{mm} \\
\texttt{expand\_dims} 
& \texttt{unsqueeze} \\
\texttt{transpose} 
& \texttt{permute} \\
\texttt{linear} 
& \texttt{dense}, \texttt{fully\_connected} \\
\texttt{batch\_norm} 
& \texttt{batch\_normalization}, \texttt{batchnorm} \\
\texttt{silu} 
& \texttt{swish} \\
\texttt{sigmoid} 
& \texttt{expit} \\
\texttt{mse\_loss} 
& \texttt{mean\_squared\_error}, \texttt{mse} \\
\texttt{kl\_divergence} 
& \texttt{kullback\_leibler\_divergence}, \texttt{kl\_div} \\
\texttt{inv} 
& \texttt{matrix\_inverse}, \texttt{inverse} \\
\bottomrule
\end{tabularx}
\end{table}

The alias mapping is used to construct candidate correspondences and does not by itself establish semantic equivalence. For APIs with different supported domains, parameter encodings, state, or default behaviors, the corresponding invocation recipe restricts execution to a shared semantic domain and translates the logical inputs into library-specific calls. A candidate correspondence is retained only after it passes execution validation.

\subsubsection{Parameter Roles}

Functionally equivalent APIs may expose the same logical inputs through different parameter names, positions, and calling conventions. \tool\ therefore normalizes framework-specific parameter names into common semantic roles. The mapping contains 111 parameter-name aliases covering 24 roles. For example, \texttt{dim}, \texttt{axis}, and \texttt{axes} are mapped to \texttt{axis}, while \texttt{input}, \texttt{x}, and \texttt{values} are mapped to a data-operand role.
Parameters that affect the tested operation, including data operands, axes, shapes, data types, dimension-retention flags, stability options, execution modes, and ordering options, are retained as shared logical inputs. Parameters with related but different encodings are translated by library-specific adapters. In contrast, 15 invocation- or execution-control parameters, such as \texttt{name}, \texttt{out}, \texttt{device}, \texttt{requires\_grad}, \texttt{jit}, and \texttt{layout} are not treated as shared logical inputs.

\begin{table}[t]
\centering
\footnotesize
\caption{Representative Parameter Roles Used by \tool.}
\label{tab:parameter-roles}
\setlength{\tabcolsep}{3pt}
\begin{tabularx}{\columnwidth}{l X}
\toprule
\textbf{Role} & \textbf{Representative Parameter Names} \\
\midrule
\texttt{data} & \texttt{input}, \texttt{x}, \texttt{values}, \texttt{lhs}, \texttt{mat1} \\
\texttt{axis} & \texttt{dim}, \texttt{axis}, \texttt{axes} \\
\texttt{shape} & \texttt{shape}, \texttt{size}, \texttt{output\_shape} \\
\texttt{dtype} & \texttt{dtype}, \texttt{output\_dtype} \\
\texttt{keepdim} & \texttt{keepdim}, \texttt{keepdims} \\
\texttt{order\_direction} & \texttt{descending}, \texttt{ascending}, \texttt{direction} \\
\texttt{training} & \texttt{training}, \texttt{train}, \texttt{is\_training} \\
\bottomrule
\end{tabularx}
\end{table}

When multiple parameters of an API share the same general role, \tool\ first uses normalized-name correspondence and then parameter position to distinguish them. Unrecognized parameters are retained using their normalized names rather than being silently discarded. The resulting roles are used to construct common logical inputs and translate them into library-specific invocations. For example, matrix-multiplication recipes restrict APIs such as \texttt{mm} to compatible two-dimensional inputs when required, while transpose and permutation APIs are invoked through operation-specific dimension mappings. Stateful operations, including dense layers and batch normalization, are compared only when their weights, statistics, and execution modes can be explicitly aligned; otherwise, the corresponding invocation is skipped.
A semantic role identifies a family of logically related parameters rather than implying direct name substitution. Repeated data operands retain their identities through normalized parameter names and positions. Parameters with inverse encodings, such as \texttt{ascending} and \texttt{descending}, are translated by library-specific adapters rather than being assigned the same Boolean value.

\subsection{Matching APIs}
\label{sec:api-matching}

\subsubsection{Constructing Candidate API Groups}

\tool\ automatically discovers public callable APIs through runtime reflection over the configured namespaces of the seven target libraries. For each API, it records the fully qualified name, callable signature, default values, normalized operation name, coarse operation category, and parameter roles. Private symbols, unavailable objects, and objects that cannot be invoked directly are excluded.

After syntactic normalization and alias resolution, APIs are initially grouped according to their normalized operation names. A coarse operation category derived from namespace and name information is used only to reject obvious semantic mismatches. A static candidate group therefore contains APIs that share the same normalized operation name, have compatible coarse categories, and originate from at least two target libraries. For example, \texttt{torch.clamp} and \texttt{tensorflow.clip\_by\_value} enter the same candidate group because both are normalized to \texttt{clip}.

This initial grouping constructs broad candidates rather than declaring functional equivalence. Name normalization and coarse filtering identify APIs that may implement the same operation, while parameter alignment and execution validation determine whether they can be invoked under common logical inputs and retained in an execution-validated API group.

\subsubsection{Aligning Input Parameters}
\label{sec:aligning-parameters}

For each candidate group, \tool\ aligns API parameters using the normalized roles defined in Table~\ref{tab:parameter-roles}. Parameters assigned to the same role represent the same logical input even when their original names, positions, or encodings differ. When multiple parameters share a role, exact normalized-name correspondence is preferred, and the original parameter order is used to resolve any remaining ambiguity.

\tool\ implements input adaptation in two layers. First, a shared input represented by NumPy arrays and Python-native values is converted into each library's native representation using library-specific tensor constructors and data-type mappings. The implementation uses \texttt{torch.tensor}, \texttt{tf.constant}, \texttt{jnp.array}, \texttt{paddle.to\_tensor}, \texttt{ms.Tensor}, and \texttt{keras.ops.convert\_to\_tensor}; Chainer directly accepts NumPy arrays.
Second, \tool\ constructs executable calls using deterministic invocation recipes indexed by canonical operation. Each recipe specifies the shared validation inputs, framework-specific parameter names and ordering, required additional arguments, and supported calling forms for that operation. These recipes handle differences such as \texttt{axis} versus \texttt{dim}, positional versus keyword arguments, and library-specific mandatory parameters. Qualified-API overrides further handle exceptional interfaces, including redirected APIs, legacy least-squares functions, CTC loss functions, and function-based control-flow operators.
For example, the same logical inputs for matrix multiplication are bound to \texttt{input} and \texttt{other} in \texttt{torch.matmul}, but to \texttt{a} and \texttt{b} in \texttt{jax.numpy.matmul}. Similarly, a shared reduction axis may be passed through \texttt{dim} in PyTorch and \texttt{axis} in TensorFlow or JAX. If no supported common invocation can be constructed, the adapter raises a \texttt{SkipCall} condition with an explicit reason, and the corresponding API pair is classified as SKIP during execution validation.

\subsubsection{Execution-Based Pair Validation}
\label{sec:pairwise-validation}

APIs with the same normalized operation name may still differ in semantics, parameter interpretation, or supported input domains. \tool\ therefore enumerates every cross-library API pair within each candidate group and validates each pair through actual execution. The same logical input is used for every supported cross-library pair in the candidate group. Boundary and non-finite values are excluded from this stage because they are explored separately during fuzzing rather than treated as evidence of an incorrect correspondence. For the canonical input, \tool\ executes both APIs and normalizes scalar, tensor, array, tuple, and list outputs. Output comparison proceeds over the normalized structure. Two outputs are non-equivalent if their structures or shapes differ. For aligned numerical outputs, \tool\ first compares their element-wise NaN masks and then their infinity masks and infinity signs. If these masks agree, identically positioned non-finite elements are excluded and the remaining finite elements are compared using the absolute tolerance $\tau=10^{-3}$. Outputs containing only identically positioned and signed non-finite values are treated as equivalent.

The validation outcomes are mutually exclusive. A pair is marked as SKIP when no supported common invocation can be constructed. For a supported invocation, it is marked as REJECT if input conversion or API execution does not complete normally, or if both APIs execute successfully but produce non-equivalent outputs. A pair is marked as PASS only when both APIs execute successfully and produce equivalent outputs. Only PASS pairs are used to construct execution-validated API groups, while SKIP and REJECT pairs do not create group edges. Each API execution is isolated so that an abnormal worker termination does not interrupt the remaining validation process.

\subsubsection{Constructing Executable API Groups}
\label{sec:group-construction}

After pairwise validation, \tool\ represents each API as a node and each PASS pair as an undirected edge. It applies union--find to the PASS edges and retains connected components that contain APIs from at least two target libraries. REJECT and SKIP pairs do not create edges.

Because pairwise PASS relationships are not necessarily transitive, a connected component may contain APIs that are linked through intermediate members but do not agree directly. \tool\ therefore reruns all APIs in each component on the same canonical validation input used during pairwise validation. A component is retained as an execution-validated API group only when all members execute successfully and produce mutually consistent outputs; otherwise, the component is excluded from subsequent fuzzing. The retained components form the execution-validated API groups used by the subsequent differential fuzzing stage. 

\subsection{Fuzzing}
\label{sec:fuzzing}

After aligning and validating the API groups, \tool\ generates common input seeds and applies variance-guided differential fuzzing to explore behavioral differences among execution-validated APIs intended to implement the same operation. Each group is tested until its configured time budget is exhausted; the specific budget is presented in the experimental setup.

\subsubsection{Input Seed Generation}

\tool\ generates input seeds using a unified parameter representation:

\begin{itemize}[leftmargin=*]
\item \textit{Tensor}: Synthetic tensors are generated as multidimensional arrays with valid ranks and shapes and are converted into each library's native tensor representation.
\item \textit{Value Scalar}: Floating-point and integer values are sampled from ranges supported by the corresponding operation.
\item \textit{Index Scalar}: Dimension and index values are constrained by the rank or size of the associated tensor.
\item \textit{Shape}: Shape-related parameters are generated as integer lists or tuples compatible with their associated tensors.
\item \textit{Boolean Flag}: Boolean parameters are assigned supported true or false values.
\end{itemize}

All inputs are first represented as NumPy arrays and Python-native values and are then translated by the library-specific adapters. This common representation ensures that all APIs in a group receive the same numerical values and logically equivalent parameter settings.

In addition to ordinary numerical values, \tool\ includes boundary and non-finite values in the input-generation and mutation pools. Representative boundary values include signed zeros, small-magnitude values, extreme finite values, and values near operation-specific limits. NaN, positive infinity, and negative infinity are injected only when the corresponding input type supports non-finite values. Boundary and non-finite injections are performed at configurable intervals.

\subsubsection{Variance-Guided Differential Fuzzing}
\label{sec:Fuzzing}

Given an execution-validated group of APIs $\{A_1,A_2,\ldots,A_n\}$ intended to implement the same operation, \tool\ supplies all APIs with the same logical input $\mathbf{x}$. Output structures and non-finite values are first processed according to Section~\ref{sec:oracles}. Only executions whose outputs have identical structures, shapes, and non-finite masks are assigned a search-guidance score. Let $\mathbf{y}_i$ denote the finite elements returned by API $A_i$, let $\bar{\mathbf{y}}$ denote their element-wise mean, and let $m$ denote the number of finite output elements. \tool\ computes the scale-normalized disagreement score as

\begin{equation}
S(\mathbf{x})=
\frac{
\frac{1}{n}\sum_{i=1}^{n}
\left\|\mathbf{y}_i-\bar{\mathbf{y}}\right\|_2^2
}{
\left\|\bar{\mathbf{y}}\right\|_2^2 + m\epsilon
},
\qquad
\bar{\mathbf{y}}=
\frac{1}{n}\sum_{i=1}^{n}\mathbf{y}_i,
\end{equation}

where $\epsilon$, a small stabilizing constant, is the machine epsilon of the normalized output type. The score is used only to guide input exploration and is not used as the final discrepancy oracle.

Starting from a valid seed, \tool\ mutates tensor elements, scalar values, indices, shapes, and Boolean parameters while preserving their types and structural constraints. Ordinary, boundary, and non-finite values are selected from the corresponding mutation pools when applicable.

A mutated input is considered improved when it increases the guidance score by at least $\Delta_{\mathrm{score}}$. When no improvement is observed for $R$ consecutive mutations, \tool\ generates a new valid seed and resumes exploration. Every execution is independently examined by the Crash and Inconsistency oracles. Executions with different output structures, shapes, or non-finite masks are processed directly by the Inconsistency oracle and are not assigned a guidance score.

\subsection{Oracles}
\label{sec:oracles}

During differential execution, \tool\ applies two oracles to flag individual executions that exhibit abnormal termination or inconsistent outputs.

\subsubsection{Crash}

The crash oracle detects unexpected crashes during API execution, including aborts, segmentation faults, assertion failures, and memory violations. If an API from one library crashes while at least one other API in the same group executes successfully under the same logical input, the test case is marked as a crash discrepancy.

\subsubsection{Inconsistency}

The inconsistency oracle performs pairwise output comparison among all successfully executed APIs in a group. An input is flagged when any pair produces different normalized structures or shapes, different element-wise NaN masks, different infinity masks, or different infinity signs at aligned positions. When the NaN and infinity masks agree, identically positioned non-finite elements are excluded and the remaining finite elements are compared using the absolute tolerance $\tau$. A finite-value mismatch that exceeds this tolerance is also reported as an output inconsistency. Thus, inconsistent NaN generation or propagation is treated as part of the unified Inconsistency oracle rather than as a separate category. The normalized disagreement score in Section~\ref{sec:Fuzzing} guides mutation selection but does not determine whether an execution is inconsistent.

\section{Evaluation}
\label{sec:evaluation}

\subsection{Research Questions}
\label{sec:research-questions}

We evaluate \tool\ by answering the following five research questions:

\noindent\textbf{RQ1:} What cross-library API groups does \tool\ construct through execution-based validation?

\noindent\textbf{RQ2:} How does the testing scope of \tool\ compare with the preliminary evaluation?

\noindent\textbf{RQ3:} How effective is \tool\ in identifying cross-library discrepancies?

\noindent\textbf{RQ4:} To what extent do the detected cases complement existing API-level differential-testing designs?

\noindent\textbf{RQ5:} How do the individual components of \tool\ contribute to its overall effectiveness?

For RQ1, we characterize the complete matching and execution-validation funnel over all seven target libraries. For RQ2, we characterize how the journal extension broadens the testing scope of \tool\ beyond the preliminary evaluation in terms of supported libraries, matched APIs, API groups, and library combinations. For RQ3, we report validated discrepancy cases across the seven libraries, including their behavioral categories, affected APIs, and developer-report status. For RQ4, we assess complementarity through CPU--GPU replay, baseline rediscovery, and TensorScope correspondence coverage. For RQ5, we evaluate execution-based matching validation, extended input generation, variance-guided fuzzing, and the per-group testing budget.

\subsection{Experimental Setup}
\label{sec:experimental-setup}

\noindent\textbf{Targeted Libraries.}
We evaluate seven numerical and deep learning libraries: PyTorch 2.12.0, TensorFlow 2.21.0, Keras 3.14.1~(evaluated using the tensorflow backend), JAX 0.10.1, MindSpore 2.9.0, PaddlePaddle 3.3.1, and Chainer 7.8.1. All matching and validation runs are performed on CPUs using fixed deterministic inputs.

%\noindent\textbf{Baseline.} We use \oldtool~\cite{xamt} as the direct baseline.

\noindent\textbf{Testing Budget.}
Unless otherwise stated, each execution-validated group is fuzzed for 60 seconds of wall-clock time, corresponding to an aggregate group-level fuzzing budget of approximately 12 hours for the 676 groups. 

\noindent\textbf{Environment.} Experiments ran on Ubuntu 24.04.3 LTS with an AMD Ryzen Threadripper PRO 7985WX CPU, 502\,GiB DDR5 memory, and 2 NVIDIA RTX 6000 Ada GPUs.

\noindent\textbf{Parameter Settings.}
After explicit comparison of output structures and non-finite masks, aligned finite output elements are compared using an absolute tolerance of $\tau=10^{-3}$, consistent with the numerical criterion used in the preliminary evaluation. A mutation is considered improved when it increases the normalized disagreement score by at least $\Delta_{\mathrm{score}}=0.001$, and the current search is restarted after $R=20$ consecutive non-improving mutations.

\subsection{Metrics}
\label{sec:metrics}

\noindent\textbf{Matched APIs and Groups.}
Following the reporting convention of the preliminary study, we report the numbers of static candidate groups, evaluated cross-library pairs, final execution-validated groups, and matched APIs contained in the final groups. 

\noindent\textbf{Code Coverage.} Code coverage has been widely adopted in software testing and recent deep learning library testing studies~\cite{gu2022muffin,deng2023largetitan,wei2022free}. Following prior deep learning library fuzzing work, we use line coverage as the coverage metric.

\noindent\textbf{Discrepancy Cases.} Following previous works~\cite{xie2022docter,deng2023largetitan}, we report the number of distinct cases detected.

\section{Result Analysis}

\subsection{RQ1: What cross-library API groups does \tool\ construct through execution-based validation?}
\label{sec:rq1}

\subsubsection{Matching and Validation Funnel}

We first examine how the matching pipeline constructs execution-validated cross-library API groups. As shown in Table~\ref{tab:matching_funnel}, name normalization, alias resolution, category filtering, and parameter-role alignment produce 765 static candidate groups. These groups represent potential correspondences rather than validated functional equivalence. \tool\ then enumerates and processes 20,202 cross-library API pairs. Among them, 11,591 are classified as PASS, 7,163 as SKIP, and 1,448 as REJECT. Only PASS pairs are retained as edges for group construction. Union--find over the PASS edges, followed by the group-level consistency check, produces 676 final execution-validated groups containing 2,563  matched APIs.
Following the reporting convention of the preliminary study, we report both the number of matched API groups and the total number of APIs contained in these groups. A group may contain APIs from multiple libraries and may include multiple equivalent API variants; therefore, the number of matched APIs is larger than the number of groups.

\begin{table}[t]
\centering
\small
\caption{Matching and Execution-Validation Funnel.}
\label{tab:matching_funnel}
\setlength{\tabcolsep}{5pt}
\renewcommand{\arraystretch}{0.85}
\begin{tabularx}{\columnwidth}{X r}
\toprule
\textbf{Stage} & \textbf{Count} \\
\midrule
Static candidate groups & 765 \\
Cross-library pairs validated & 20,202 \\
\quad PASS & 11,591 \\
\quad SKIP & 7,163 \\
\quad REJECT & 1,448 \\
Final execution-validated groups & 676 \\
Matched APIs & 2,563  \\
\bottomrule
\end{tabularx}
\end{table}

\subsubsection{Matched APIs Across Libraries}

Table~\ref{tab:matched_group_stats} reports the number of APIs from each library retained in the final execution-validated groups. PyTorch contributes 480 matched APIs, followed by MindSpore with 441, PaddlePaddle with 431, JAX with 379, Keras with 331, TensorFlow with 293, and Chainer with 208. These counts follow the same matched-API reporting convention as the preliminary evaluation. Because the matching process does not require a designated reference library, a final group may contain APIs from any combination of at least two target libraries.

\begin{table}[t]
\centering
\footnotesize
\caption{Matched APIs Across the Seven Libraries.}
\label{tab:matched_group_stats}
\setlength{\tabcolsep}{2.2pt}
\renewcommand{\arraystretch}{0.85}
\begin{tabular}{lrrrrrrr}
\toprule
 & \textbf{PyTorch} & \textbf{MindSpore} & \textbf{Paddle} &
\textbf{JAX} & \textbf{Keras} & \textbf{TensorFlow} &
\textbf{Chainer} \\
\midrule
\#APIs
& 480 & 441 & 431 & 379 & 331 & 293 & 208 \\
\bottomrule
\end{tabular}
\end{table}

\subsubsection{Composition of Execution-Validated Groups}

\begin{figure}[t]
  \centering
  \includegraphics[width=\linewidth]{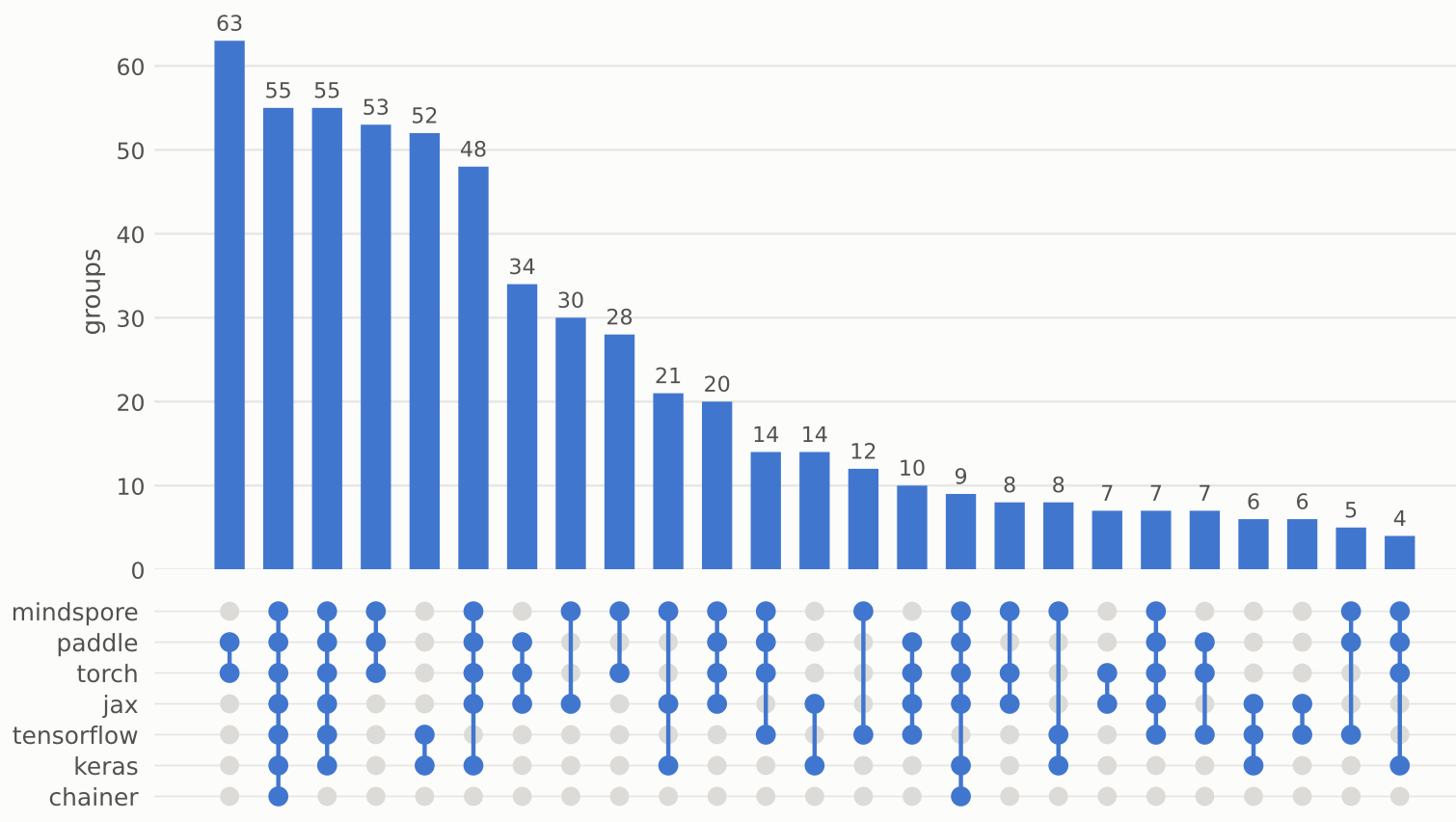}
  \caption{Distribution of API groups across library combinations.}
  \label{fig:api_upset_plot}
\end{figure}

Figure~\ref{fig:api_upset_plot} shows the 25 most frequent library combinations among the 74 combinations observed in the 676 execution-validated groups. The seven rows represent MindSpore, PaddlePaddle, PyTorch, JAX, TensorFlow, Keras, and Chainer. Each column represents one exact combination of these libraries: blue dots indicate the participating libraries, vertical lines connect the libraries belonging to the same combination, and the bar above reports the number of API groups with that combination. 
The final groups span 74 distinct combinations of the seven target libraries. Among the 676 groups, 232 contain APIs from two libraries, while 444 contain APIs from three or more libraries. Thus, 65.7\% of the final groups include at least three library implementations. Multi-library groups provide additional implementations for differential comparison, whereas two-library groups retain operations available in only a limited subset of the target libraries.

\subsubsection{Effect of Execution Validation}

Pairwise validation separates the 20,202 candidate pairs into three outcomes. PASS pairs provide behavioral evidence for group construction, while SKIP pairs indicate that no supported common invocation can be constructed. The 1,448 REJECT outcomes comprise 1,264 invocations that do not complete normally and 184 successful executions that produce non-equivalent outputs on the canonical validation input. We separately examine the 184 output mismatches using the same finding-retention procedure applied to fuzzing candidates. None provides sufficient evidence for retention because the differences arise from unsuitable candidate correspondences, incompatible supported domains, parameter or default semantics, or acceptable numerical variation.
The PASS relationships are used to construct connected components, which are then examined by the group-level consistency check. This process produces the 676 final execution-validated groups used for subsequent fuzzing. The group-level check serves as a safeguard against non-transitive pairwise agreement. As with any input-based validation, these checks do not establish semantic equivalence over the complete input domain.

We also manually audit 50 PASS, 50 REJECT, and 50 SKIP pairs. The audit finds that 100\% of the PASS pairs are semantically comparable and that 96\% of the REJECT pairs are correctly excluded. Among the audited SKIP pairs, 62\% are caused by unsupported invocation recipes, 30\% by the absence of a common input domain, and 8\% by other interface or environment limitations.

\noindent\textbf{Answer to RQ1.} Across seven libraries, \tool\ constructs 676 cross-library groups that pass the configured pairwise and group-level execution checks. The final groups contain 2,563  matched APIs and span 74 library combinations. Pairwise validation produces 11,591 PASS, 7,163 SKIP, and 1,448 REJECT outcomes, and 444 groups contain APIs from at least three libraries.

\subsection{RQ2: How does the testing scope of \tool\ compare with the preliminary evaluation?}
\label{sec:rq2}

\subsubsection{Expansion of the Testing Scope}

We compare the API-level testing scope of the preliminary and extended evaluations of \tool. As shown in Table~\ref{tab:rq2_api_coverage}, the preliminary evaluation used PyTorch as the reference library and constructed 238 cross-library groups containing 839 matched APIs across five libraries. The journal extension removes the designated-reference-library requirement, expands the target set to seven libraries, and constructs 676 execution-validated groups containing 2,563  matched APIs. These groups span 74 distinct library combinations.

\begin{table}[t]
\centering
\small
\caption{Expansion of the API-Level Testing Scope of \tool.}
\label{tab:rq2_api_coverage}
\setlength{\tabcolsep}{10pt}
\renewcommand{\arraystretch}{0.85}
\begin{tabular}{lrrr}
\toprule
\textbf{Evaluation} &
\textbf{Libraries} &
\textbf{Matched APIs} &
\textbf{Groups} \\
\midrule
ISSRE & 5 & 839 & 238 \\
Extended      & 7 & 2,563  & 676 \\
\bottomrule
\end{tabular}
\end{table}

These figures describe the expanded evaluation scope of the same approach rather than a comparison between two independent techniques.

\subsubsection{Code Coverage}

To preserve comparability with the preliminary evaluation, we report line coverage under the same aggregate testing budget of 12 hours. Table~\ref{tab:rq2_code_coverage} presents the results for the five libraries shared by both evaluations, together with the MindSpore and PaddlePaddle results included only in the journal extension.

For the five shared libraries, the extended evaluation reports consistently higher line coverage. PyTorch coverage increases from 19.97\% to 25.18\%, TensorFlow from 22.19\% to 26.74\%, Keras from 24.22\% to 29.63\%, JAX from 28.98\% to 33.91\%, and Chainer from 25.45\% to 28.42\%. The corresponding increases range from 2.97 to 5.41 percentage points. Across the five libraries, the average line coverage increases from 24.16\% in the preliminary evaluation to 28.78\% in the extended evaluation, an absolute improvement of 4.62 percentage points. The journal extension additionally reports 18.76\% coverage for MindSpore and 21.53\% for PaddlePaddle.

The preliminary and extended evaluations retain the same variance-guided fuzzing strategy and use the same total testing budget of 12 hours. The reported coverage improvement should therefore not be interpreted as evidence of a newly introduced search algorithm. Because the two evaluations use different library versions and API universes, the comparison reflects the combined effects of the expanded matching scope, input domain, and evaluated API set rather than the isolated contribution of any single component.

\begin{table}[t]
\centering
\small
\caption{Line Coverage under the Same 12-Hour Total Testing Budget.}
\label{tab:rq2_code_coverage}
\setlength{\tabcolsep}{13pt}
\renewcommand{\arraystretch}{0.85}
\begin{tabular}{lrrr}
\toprule
\textbf{Library} &
\textbf{ISSRE} &
\textbf{Extended} &
\textbf{$\Delta$} \\
\midrule
PyTorch      & 19.97\% & 25.18\% & +5.21\% \\
TensorFlow   & 22.19\% & 26.74\% & +4.55\% \\
Keras        & 24.22\% & 29.63\% & +5.41\% \\
JAX          & 28.98\% & 33.91\% & +4.93\% \\
Chainer      & 25.45\% & 28.42\% & +2.97\% \\
\midrule
MindSpore    & --      & 18.76\% & -- \\
PaddlePaddle & --      & 21.53\% & -- \\
\bottomrule
\end{tabular}
\end{table}

\noindent\textbf{Answer to RQ2.}
Under the same total testing budget of 12 hours, the journal extension broadens the evaluation of \tool\ from five to seven libraries, increases the number of matched APIs from 839 to 2,563 , and increases the number of cross-library groups from 238 to 676. The resulting groups span 74 library combinations and no longer depend on a designated reference library. For the five shared libraries, the average line coverage increases from 24.16\% in the preliminary evaluation to 28.78\% in the extended evaluation, corresponding to an absolute improvement of 4.62 percentage points.

\subsection{RQ3: How effective is \tool\ in identifying cross-library discrepancies?}
\label{sec:rq3}

\subsubsection{Discrepancy Categories}

We classify the independently reproduced discrepancy cases by their observed behavior: abnormal process termination or output inconsistency. As shown in Table~\ref{tab:rq3_categories}, four cases trigger abnormal process termination and 68 involve output inconsistencies, including differences in output structures, shapes, non-finite masks, infinity signs, or finite values.

The four Crash cases were discovered during fuzzing and represent abnormal worker termination rather than handled API exceptions. The Inconsistency cases involve materially different outputs produced by APIs retained in the same execution-validated group.

\begin{table}[t]
\centering
\small
\caption{Validated Discrepancy Cases by Oracle Category.}
\label{tab:rq3_categories}
\setlength{\tabcolsep}{2.5pt}
\renewcommand{\arraystretch}{0.85}
\begin{tabularx}{\columnwidth}{
    l
    >{\centering\arraybackslash}X
    >{\centering\arraybackslash}X
    >{\centering\arraybackslash}X}
\toprule
\textbf{Category} &
\textbf{Crash} &
\textbf{Inconsistency} &
\textbf{Total} \\
\midrule
\textbf{\#Cases} & 4 & 68 & 72 \\
\bottomrule
\end{tabularx}
\end{table}

Each retained case is independently reproduced before inclusion in the final results. We examine its triggering input, API semantics, and observed outputs using official API specifications and, when applicable, numerical references such as NumPy or SciPy. When no independent numerical reference is available, we report the reproducible behavioral discrepancy without treating majority behavior alone as definitive ground truth.

\subsubsection{Developer Reporting Status}

We submitted 72 developer reports for the validated discrepancies. Table~\ref{tab:rq3_issue_status} summarizes their current status by library. Fixed reports are included in the Confirmed column because the corresponding issues have been accepted and addressed by the developers. The replication package provides the complete list of developer-report links~\footnote{\url{https://github.com/beanduan22/SupXamt/tree/main}}.

Overall, 25 reports have been confirmed, including 23 that have already been fixed. Another 41 reports remain pending, while six have other dispositions. Pending and other reports are not treated as maintainer-confirmed defects. In particular, all 27 MindSpore reports remain pending and are therefore not counted as confirmed defects. Every retained case has nevertheless been independently reproduced and examined as described above.

\begin{table}[t]
\centering
\small
\caption{Current Developer-Reporting Status of the Validated Discrepancies.}
\label{tab:rq3_issue_status}
\setlength{\tabcolsep}{4pt}
\renewcommand{\arraystretch}{0.85}
\begin{tabular}{lrrrrr}
\toprule
\textbf{Library} &
\textbf{Reports} &
\textbf{Confirmed} &
\textbf{Fixed} &
\textbf{Pending} &
\textbf{Other} \\
\midrule
PyTorch      & 7  & 5 & 5 & 0  & 2 \\
TensorFlow   & 5  & 5 & 3 & 0  & 0 \\
Keras        & 7  & 7 & 7 & 0  & 0 \\
JAX          & 12 & 7 & 7 & 1  & 4 \\
PaddlePaddle & 12 & 1 & 1 & 11 & 0 \\
Chainer      & 2  & 0 & 0 & 2  & 0 \\
MindSpore    & 27 & 0 & 0 & 27 & 0 \\
\midrule
Total        & 72 & 25 & 23 & 41 & 6 \\
\bottomrule
\end{tabular}
\end{table}

Output inconsistency is the dominant category, accounting for 68 of the 72 cases. These cases include structural, non-finite-mask, infinity-sign, and finite-value disagreements. The four Crash cases show that the testing process can also expose abnormal process termination. Developer responses currently provide external confirmation for 25 reports, although most recently submitted reports remain pending.

\subsubsection{Representative Fixed Cases}

Here, we present three representative discrepancies that were fixed after reporting. They illustrate different implementation problems, including incorrect boundary handling, inconsistent comparison semantics, and unintended propagation of non-finite values.

\begin{figure}[t!]
\centering
\begin{minipage}{\linewidth}
\lstset{style=mystyle}
\begin{lstlisting}[language=Python, label=bug_erfinv]
x = np.nextafter(
    np.float32(1.0),
    np.float32(0.0)
)
tensorflow_result = tf.math.erfinv(x)
\end{lstlisting}

\lstset{style=mystyle_1}
\begin{lstlisting}[language=Python, caption=Incorrect Boundary Result in TensorFlow, label=bug_erfinv_out]
output:
TensorFlow result: inf
Reference result:  3.8325071
\end{lstlisting}
\end{minipage}
\label{fig:bug-erfinv-example}
\end{figure}

The discrepancy in Listing~\ref{bug_erfinv_out} is triggered by the largest representable \texttt{float32} value below 1. TensorFlow returns \texttt{inf}, whereas the reference returns a finite result. This case reveals incorrect numerical handling near the boundary of the valid input domain. The issue was reported to TensorFlow and has been fixed.

\begin{figure}[t!]
\centering
\begin{minipage}{\linewidth}
\lstset{style=mystyle}
\begin{lstlisting}[language=Python, label=bug_isclose]
x = np.array(
    [np.nan, np.inf, -np.inf],
    dtype=np.float32
)
keras_result = keras.ops.isclose(x, x)
\end{lstlisting}

\lstset{style=mystyle_1}
\begin{lstlisting}[language=Python, caption=Incorrect Infinity Comparison in Keras, label=bug_isclose_out]
output:
Keras result: [False False False]
Reference result: [False  True  True]
\end{lstlisting}
\end{minipage}
\label{fig:bug-isclose-example}
\end{figure}

The discrepancy in Listing~\ref{bug_isclose_out} shows that Keras incorrectly treats identical positive and negative infinities as unequal. Although \texttt{NaN} is expected to remain unequal to itself, matching infinity values should be considered equal. The issue concerns comparison semantics rather than numerical tolerance and has been fixed by the developers.

\begin{figure}[t!]
\centering
\begin{minipage}{\linewidth}
\lstset{style=mystyle}
\begin{lstlisting}[language=Python, label=bug_toeplitz]
c = np.array([np.nan, np.inf, -0.0])
r = np.array([np.nan, 2.0, 3.0])

jax_result   = jax.scipy.linalg.toeplitz(c, r)
\end{lstlisting}

\lstset{style=mystyle_1}
\begin{lstlisting}[language=Python, caption=Incorrect NaN Propagation in JAX, label=bug_toeplitz_out]
output:
JAX result:
[[nan nan nan] [nan nan nan] [nan nan nan]]
Reference result:
[[nan  2.0  3.0] [inf  nan  2.0] [-0.0 inf  nan]]
\end{lstlisting}
\end{minipage}
\label{fig:bug-toeplitz-example}
\end{figure}

The discrepancy in Listing~\ref{bug_toeplitz_out} shows that JAX propagates a local \texttt{NaN} value across the entire Toeplitz matrix. In contrast, the reference preserves the expected matrix structure and confines non-finite values to positions determined by the input vectors. This incorrect propagation behavior was reported and subsequently fixed.

These fixed cases illustrate three distinct forms of incorrect behavior: a finite boundary input producing infinity, incorrect Boolean comparison semantics for infinity values, and excessive propagation of \texttt{NaN} across an output matrix. They demonstrate that the detected discrepancies are not limited to small numerical deviations, but also include clear violations of expected API behavior.

\noindent\textbf{Answer to RQ3.}
\tool\ identifies 72 validated discrepancy cases, including four Crash cases and 68 output inconsistencies. Among the 72 submitted reports, 25 have been confirmed, including 23 that have been fixed; 41 remain pending and six have other dispositions.

\subsection{RQ4: To what extent do the detected cases complement existing API-level differential-testing designs?}
\label{sec:rq4}

We examine complementarity at three levels. First, fixed-input backend replay determines whether the triggering inputs already provide a direct CPU--GPU disagreement signal. Second, method-level rediscovery evaluates whether FreeFuzz and DeepREL can expose the same underlying defects using their own input-generation strategies and native oracle sets. Third, a correspondence-scope audit determines whether TensorScope can represent the cross-library relations witnessing the detected cases. These analyses characterize different forms of complementarity and do not establish that a defect is unreachable under every possible input or configuration.

\subsubsection{Oracle-Level Comparison with CPU--GPU Differential Testing}

We replay the fixed triggering inputs for the 60 cases included in the existing backend-replay dataset on the CPU and GPU implementations of the affected APIs whenever both backends support the input. These cases are a subset of the unified Inconsistency category. Output structures and non-finite values are processed using the same rules as the main Inconsistency oracle.
All 60 replayed cases produce a cross-library disagreement. Among them, 50 support both CPU and GPU execution, but none of the fixed inputs produces an observable backend disagreement. The remaining 10 cases involve MindSpore APIs for which a compatible GPU execution is unavailable. This fixed-input replay shows that the evaluated inputs provide no signal to a CPU--GPU differential oracle; it does not establish that a complete backend-differential testing technique using its own input-generation strategy could never expose the same defects through other inputs.

\subsubsection{Method-Level Rediscovery}

We run FreeFuzz and DeepREL on the affected APIs supported by their original workflows, using their own input-generation strategies and native oracle sets. A case is considered rediscovered when a baseline-generated test exposes the same underlying defect, even through a different input or observable symptom. Two authors independently examine each candidate; unresolved root-cause attributions are classified as inconclusive.

\begin{table}[t]
\centering
\footnotesize
\caption{Defect-Level Rediscovery by Existing Baselines.}
\label{tab:baseline_rediscovery}
\setlength{\tabcolsep}{3pt}
\renewcommand{\arraystretch}{0.9}
\begin{tabular}{lrrrr}
\toprule
\textbf{Baseline} &
\textbf{Applicable} &
\textbf{Rediscovered} &
\textbf{Not Rediscovered} &
\textbf{Inconclusive} \\
\midrule
FreeFuzz & 15 & 2 & 12 & 1 \\
DeepREL  & 11 & 1 & 9  & 1 \\
\bottomrule
\end{tabular}
\end{table}

As shown in Table~\ref{tab:baseline_rediscovery}, FreeFuzz rediscovered 2 of the 15 applicable defects, while DeepREL rediscovered 1 of the 11 applicable defects. The remaining applicable defects were not rediscovered under the evaluated versions and testing budgets, except for one inconclusive attribution for each baseline. Cases outside their supported frameworks, APIs, or relational-oracle scopes are treated as not applicable rather than unsuccessful rediscovery. These results indicate that \tool\ complements the evaluated baselines within their supported testing scopes.

\subsubsection{Correspondence-Scope Comparison with TensorScope}

TensorScope derives directed counterpart relations from the registries and conversion handlers of eight model converters. A source API may correspond to either one destination API or an ordered composition of multiple destination APIs. Its matching scope is therefore bounded by the frameworks and conversion rules represented in the examined converter ecosystem. In contrast, \tool\ does not require a converter path and constructs multi-library groups that are checked through pairwise and group-level execution before fuzzing. TensorScope covers TensorFlow, TensorFlow Lite, ONNX Runtime, MindSpore, PyTorch, and PaddlePaddle; JAX and Chainer are outside its framework matrix, and no explicit Keras API-side correspondence registry is included.

We audit the 72 independently reproduced discrepancy cases against the converter registries and handlers identified by TensorScope. For each case, we identify the API pair or pairs that actually witness the disagreement and check whether the required relation is represented directly or through an intermediate ONNX path. A case is \textit{fully represented} when at least one witnessing correspondence, including its required operation and parameter relation, is available; \textit{partially represented} when relevant operations appear but the complete witnessing correspondence is unavailable; and \textit{not represented} when none of its witnessing correspondences can be constructed from the examined conversion rules.

\begin{table}[t]
\centering
\footnotesize
\caption{TensorScope Representation of the 72 Cases.}
\label{tab:tensorscope_scope}
\setlength{\tabcolsep}{10pt}
\renewcommand{\arraystretch}{0.9}
\begin{tabular}{lrr}
\toprule
\textbf{Representation} &
\textbf{Cases} &
\textbf{Percentage} \\
\midrule
Fully represented     & 7  & 9.7\% \\
Partially represented & 22 & 30.6\% \\
Not represented       & 43 & 59.7\% \\
\midrule
Total                  & 72 & 100.0\% \\
\bottomrule
\end{tabular}
\end{table}

As shown in Table~\ref{tab:tensorscope_scope}, only 7 of the 72 cases (9.7\%) are fully represented in TensorScope's converter-derived correspondence scope. Another 22 cases (30.6\%) are partially represented, while 43 cases (59.7\%) involve witnessing correspondences that cannot be constructed from the examined converter registries and handlers. Overall, 65 cases (90.3\%) are not fully represented. This substantial scope difference arises from API relations involving libraries outside TensorScope's framework matrix, explicit Keras API-side interfaces, and operations or parameter mappings absent from the examined conversion rules.
This result characterizes matching applicability rather than end-to-end detection effectiveness. A fully represented case may still require a triggering input that TensorScope does not generate, whereas a not-represented case cannot be tested through the same cross-framework relation without extending its converter-derived correspondence scope.

\noindent\textbf{Answer to RQ4.}
The analysis shows complementarity at three levels. None of the fixed triggering inputs of the 50 CPU--GPU-applicable cases in the backend-replay subset produces a backend-disagreement signal. FreeFuzz and DeepREL rediscover 2 of 15 and 1 of 11 applicable defects, respectively, under the evaluated configurations. In addition, only 7 of the 72 cases are fully represented in TensorScope's converter-derived correspondence scope, while 22 are partially represented and 43 are not represented.

\subsection{RQ5: How do the individual components of \tool\ contribute to its overall effectiveness?}
\label{sec:rq5}

\begin{table}[t]
\centering
\footnotesize
\caption{Component Analysis and Tolerance Sensitivity.}
\label{tab:rq5_components}
\setlength{\tabcolsep}{5pt}
\renewcommand{\arraystretch}{0.9}
\begin{tabular}{lccc}
\toprule
\multicolumn{4}{c}{\textbf{(a) Execution Validation}} \\
\midrule
\textbf{Configuration} &
\textbf{Groups} &
\textbf{Raw} &
\textbf{Raw/100 Groups} \\
\midrule
Full \tool                   & 676 & 109 & 16.1 \\
w/o Execution Validation    & 765 & 194 & 25.4 \\
\midrule
\multicolumn{4}{c}{\textbf{(b) Input Generation}} \\
\midrule
\textbf{Configuration} &
\textbf{Avg. Coverage} &
\textbf{Raw} &
\textbf{Reference Cases} \\
\midrule
Full \tool                  & 26.31\% & 109 & 72/72 \\
w/o Extended Inputs        & 25.63\% & 78  & 50/72 \\
Random Mutation            & 24.47\% & 73  & 47/72 \\
\midrule
\multicolumn{4}{c}{\textbf{(c) Oracle Tolerance}} \\
\midrule
\textbf{Tolerance} &
\multicolumn{2}{c}{\textbf{Inconsistency Cases}} &
\textbf{Change} \\
\midrule
$10^{-4}$ & \multicolumn{2}{c}{68} & 0 \\
$10^{-3}$ & \multicolumn{2}{c}{68} & -- \\
$10^{-2}$ & \multicolumn{2}{c}{60} & -8 \\
\bottomrule
\end{tabular}
\end{table}

For the component analyses in RQ5, each tested API group receives the same wall-clock budget of 60 seconds. This per-group budget controls the testing opportunity assigned to each group across configurations. Because different configurations may contain different numbers of groups, their aggregate runtimes may differ.

\subsubsection{Contribution of Execution-Based Matching Validation}

We examine whether execution-based validation reduces unsuitable API correspondences supplied to fuzzing. The full configuration applies pairwise execution validation and the group-level consistency check, producing 676 execution-validated groups. The ablated configuration skips both validation stages and directly tests all 765 static candidate groups. Both configurations use the same extended input generation, variance-guided fuzzing, test oracles, and per-group budget of 60 seconds.
As shown in Table~\ref{tab:rq5_components}, the full configuration produces 109 raw discrepancy candidates, corresponding to 16.1 candidates per 100 tested groups. Without execution validation, the number increases to 194, or 25.4 candidates per 100 groups. Execution validation therefore reduces the density of downstream candidates requiring reproduction and examination by filtering unsuitable correspondences before fuzzing. This analysis evaluates its candidate-filtering role rather than its effect on the final retained-case set.

\subsubsection{Contribution of Extended Input Generation}

We evaluate the contribution of boundary and non-finite inputs by retaining the same execution-validated groups, variance-guided search, mutation operators, test oracles, initial seeds, and 60-second per-group budget, while restricting input generation and mutation to ordinary finite values.
We use the 72 discrepancy cases identified by the full configuration as a fixed reference set and report how many of their triggering conditions are reached under each ablated configuration. As shown in Table~\ref{tab:rq5_components}, removing extended inputs decreases the average line coverage from 26.31\% to 25.63\% and reaches 50 of the 72 reference cases. Among the 22 cases not reached, 20 require boundary or non-finite triggering values, while the remaining two are not reached within the testing budget. These results show that extended inputs primarily improve the exploration of boundary and non-finite behaviors.

\subsubsection{Contribution of Variance-Guided Fuzzing}

We compare variance-guided fuzzing with random mutation. The Random Mutation configuration uses the same execution-validated groups, initial seeds, complete input domain, mutation operators, test oracles, and 60-second per-group budget. It differs only in that structurally valid mutations are explored without prioritization based on the normalized disagreement score.
As shown in Table~\ref{tab:rq5_components}, random mutation achieves an average line coverage of 24.47\% and reaches 47 of the 72 reference cases, compared with 26.31\% and 72 cases for the full configuration. These results indicate that the disagreement-guided strategy improves both code exploration and the ability to reach the triggering conditions of the reference cases.

\subsubsection{Sensitivity to the Oracle Tolerance}

We replay the fixed triggering inputs of the 68 Inconsistency cases using the absolute tolerance $\tau\in\{10^{-4},10^{-3},10^{-2}\}$, where $10^{-3}$ is the default value used in the main evaluation. This analysis changes only the finite-output comparison threshold and does not rerun the fuzzing process. Mask-level and infinity-sign discrepancies are independent of the finite-value tolerance. As shown in Table~\ref{tab:rq5_components}, all 68 cases remain classified as inconsistencies at $10^{-4}$ and the default $10^{-3}$, while 60 cases remain at $10^{-2}$. Thus, 88.2\% of the cases remain detectable under the most permissive evaluated tolerance, indicating that most findings are not artifacts of a narrowly selected threshold.

\begin{figure}[t]
\centering
\begin{tikzpicture}
\begin{axis}[
    width=0.8\columnwidth,
    height=4cm,
    xlabel={Budget per Group (s)},
    ylabel={Line Coverage (\%)},
    xmin=25, xmax=125,
    ymin=25.6, ymax=26.6,
    xtick={30,60,120},
    ytick={25.75,26.00,26.25,26.50},
    grid=both,
    major grid style={dashed,gray!30},
    minor grid style={dotted,gray!20},
    enlargelimits=false,
    tick label style={font=\footnotesize},
    label style={font=\footnotesize},
]
\addplot[
    color=black,
    mark=*,
    line width=1pt,
    error bars/.cd,
        y dir=both,
        y explicit
] coordinates {
(30,25.75)
(60,26.31)
(120,26.41)
};
\end{axis}
\end{tikzpicture}
\caption{Sensitivity of total testing budget.}
\label{fig:rq5_budget_cov}
\end{figure}
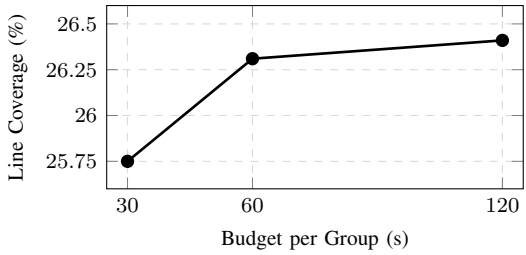

\subsubsection{Sensitivity to the Testing Budget}

We compare three per-group testing budgets: 30, 60, and 120 seconds. Fig.~\ref{fig:rq5_budget_cov} shows that the line coverage increases from 25.75\% at 30 seconds to 26.31\% at 60 seconds and 26.41\% at 120 seconds. Increasing the budget from 30 to 60 seconds adds 0.56 percentage points, whereas increasing it from 60 to 120 seconds adds only 0.10 percentage points. We therefore use 60 seconds per group as the default budget.

\noindent\textbf{Answer to RQ5.}
Execution validation reduces the density of raw discrepancy candidates from 25.4 to 16.1 per 100 tested groups. Under the same 60-second per-group budget, the full configuration achieves an average line coverage of 26.31\%, compared with 25.63\% without extended inputs and 24.47\% under random mutation. Without extended inputs and under random mutation, 50 and 47 of the 72 reference cases are reached, respectively. The budget analysis shows that coverage largely stabilizes beyond 60 seconds per group.

\section{Related Work}

\subsection{Testing Deep Learning Libraries}

The reliability of deep learning libraries has been studied through testing, program analysis, differential execution, and fuzzing~\cite{yang2019synergistic,yang2019advances,yang2007algebraic,yi2022feedback,mahmud2022acid,yi2020summary,wu2022evaluating}. Existing testing techniques target different abstraction levels, including complete deep learning models~\cite{christou2023ivysyn}, compiler and backend implementations~\cite{xiao2022metamorphic}, computational graphs~\cite{wang2022eagle}, and individual library APIs~\cite{deng2022fuzzing}. Our work focuses on API-level testing, but is also related to model-level and cross-framework differential testing.

Model-level testing generates or mutates complete neural networks and observes their execution across libraries, backends, or configurations. CRADLE~\cite{pham2019cradle} detects inconsistencies by executing existing models through different deep learning libraries. LEMON~\cite{wang2020deep} and AUDEE~\cite{guo2020audee} improve model diversity by mutating seed models and their inputs, while Muffin~\cite{gu2022muffin} generates models through a top-down construction process. NNSmith~\cite{liu2023nnsmith} combines symbolic constraint solving with gradient-guided search to synthesize executable models.

These techniques exercise interactions among multiple operators and can expose model-level failures. However, a generated model must maintain compatible tensor shapes, types, and operator dependencies, which may limit direct exploration of the input domains of individual APIs. In contrast, API-level testing can target a specific operation and systematically vary its arguments without embedding it in a complete model.

\subsection{API-Level Fuzzing}

API-level fuzzing directly generates inputs for individual library APIs. FreeFuzz~\cite{wei2022free} mines valid API invocations from open-source programs and mutates the observed inputs to expose unexpected behavior. DocTer~\cite{xie2022docter} extracts input constraints from API documentation and generates constraint-aware tests, particularly for crash detection. DeepREL~\cite{deng2022fuzzing} identifies related APIs within a library and uses relational and automatic-differentiation properties as test oracles.

Recent approaches further employ large language models to generate executable test programs. TitanFuzz~\cite{deng2023largetitan} uses a language model to generate and mutate deep learning programs while preserving API dependencies and input constraints. FuzzGPT~\cite{deng2023large} reuses patterns from historical bug-triggering programs to generate tests that exercise uncommon behaviors~\cite{chen2019history,holler2012fuzzing}. These approaches improve input and program diversity, but generally test APIs using intra-library properties, exceptions, backend differences, or execution relations rather than systematically comparing implemented equivalents across multiple libraries.

Many deep learning library testing techniques use intra-library differential oracles, including comparisons between CPU and GPU executions. Such comparisons are effective when a defect manifests differently across hardware backends, but provide no disagreement signal when an implementation produces the same divergent result on both devices. In RQ4, we replay the fixed \tool\ triggering inputs under a CPU--GPU differential oracle. None of the 50 applicable output inconsistencies produces an observable backend disagreement, showing that these cross-library discrepancies are not exposed by CPU--GPU comparison under the evaluated configurations.

\subsection{Cross-Framework Differential Testing}

Cross-framework differential testing compares independently implemented operations that are expected to provide equivalent functionality. It complements intra-library testing by providing additional implementations as behavioral references. The central challenge is constructing comparable API groups and translating a shared logical input into compatible library-specific invocations.
TensorScope~\cite{deng2023differential} extracts directed counterpart relations and parameter mappings from model-converter registries and handlers. A source API may correspond to either one destination API or an ordered composition of multiple destination APIs. Its correspondence scope is therefore bounded by the frameworks and conversion rules represented in the examined converter ecosystem. In contrast, \tool\ does not require a converter path and retains multi-library correspondences after pairwise and group-level execution validation. RQ4 quantitatively examines the extent to which the witnessing relations of the \tool\ cases are represented by TensorScope's converter-derived scope.

\section{Threats to Validity}

\noindent\textbf{Internal Validity.}
Errors in API aliases, parameter-role mappings, invocation recipes, or library-specific adapters may incorrectly include or exclude API correspondences. We mitigate this threat by checking the curated mappings against official documentation, testing representative operation categories, and validating supported correspondences through execution on shared logical inputs. Nevertheless, execution validation is input-dependent: each supported pair is checked using one deterministic operation-specific ordinary input and is therefore not proven equivalent over the complete input domain. Fuzzing results may also be affected by randomness and library-level nondeterminism. In a post-hoc audit, we re-executed 100 sampled groups on a second ordinary validation input: 100 remained mutually consistent. We fix random seeds and use deterministic CPU execution whenever possible, but a finite testing budget cannot guarantee that every reachable discrepancy is exposed. The comparison in RQ2 uses the same aggregate 12-hour budget as the preliminary evaluation, whereas the component analyses in RQ5 use the same 60-second per-group budget. Finally, the cross-library oracles identify observable disagreements but do not independently determine which implementation is incorrect. 

\noindent\textbf{External Validity.}
Our evaluation covers seven libraries and may not generalize to other libraries, versions, namespaces, hardware backends, or programming languages. The evaluated API set is limited to public callable objects in the configured namespaces, and some APIs cannot be tested because no cross-library counterpart or supported common invocation can be constructed. The findings are also specific to the evaluated software versions and environment. In addition, Keras is evaluated using the TensorFlow backend, so the corresponding implementations are not fully independent. Extending the invocation recipes, validation inputs, target libraries, Keras backends, and hardware environments would improve generalizability.

\section{Conclusion}

This paper presents \tool, a cross-framework differential fuzzing approach for deep learning library APIs. \tool\ constructs candidate API correspondences through explicit API aliases and parameter-role normalization, retains supported correspondences after pairwise and group-level execution validation, and applies variance-guided fuzzing over ordinary, boundary, and non-finite inputs. Across seven libraries, \tool\ constructs 676 execution-validated groups containing 2,563 API members. The evaluation retains 72 distinct discrepancy cases, including four involving abnormal process termination and 68 involving output inconsistencies. Among the 72 submitted reports, 25 have been confirmed, including 23 that have been fixed. The results demonstrate that cross-library comparison exposes API-level discrepancies that provide no CPU--GPU disagreement signal under the evaluated fixed inputs and configurations.

\balance
\bibliographystyle{IEEEtran}
\bibliography{main}

@inproceedings{deng2023differential,
  title={Differential testing of cross deep learning framework $\{$APIs$\}$: Revealing inconsistencies and vulnerabilities},
  author={Deng, Zizhuang and Meng, Guozhu and Chen, Kai and Liu, Tong and Xiang, Lu and Chen, Chunyang},
  booktitle={32nd USENIX Security Symposium (USENIX Security 23)},
  pages={7393--7410},
  year={2023}
}

@inproceedings{wu2022evaluating,
  title={Evaluating and improving neural program-smoothing-based fuzzing},
  author={Wu, Mingyuan and Jiang, Ling and Xiang, Jiahong and Zhang, Yuqun and Yang, Guowei and Ma, Huixin and Nie, Sen and Wu, Shi and Cui, Heming and Zhang, Lingming},
  booktitle={Proceedings of the 44th International Conference on Software Engineering},
  pages={847--858},
  year={2022}
}

@article{yang2019synergistic,
  title={A synergistic approach to improving symbolic execution using test ranges},
  author={Yang, Guowei and Qiu, Rui and Khurshid, Sarfraz and P{\u{a}}s{\u{a}}reanu, Corina S and Wen, Junye},
  journal={Innovations in Systems and Software Engineering},
  volume={15},
  pages={325--342},
  year={2019},
  publisher={Springer}
}

@article{yang2019advances,
  title={Advances in symbolic execution},
  author={Yang, Guowei and Filieri, Antonio and Borges, Mateus and Clun, Donato and Wen, Junye},
  journal={Advances in Computers},
  volume={113},
  pages={225--287},
  year={2019},
  publisher={Elsevier}
}

@inproceedings{yang2007algebraic,
  title={An algebraic approach for managing inconsistencies in software processes},
  author={Yang, Qiusong and Li, Mingshu and Wang, Qing and Yang, Guowei and Zhai, Jian and Li, Juan and Hou, Lishan and Yang, Yun},
  booktitle={International Conference on Software Process},
  pages={121--133},
  year={2007},
  organization={Springer}
}

@inproceedings{yi2022feedback,
  title={Feedback-driven incremental symbolic execution},
  author={Yi, Qiuping and Yang, Guowei},
  booktitle={2022 IEEE 33rd International Symposium on Software Reliability Engineering (ISSRE)},
  pages={505--516},
  year={2022},
  organization={IEEE}
}

@inproceedings{mahmud2022acid,
  title={ACID: an API compatibility issue detector for Android apps},
  author={Mahmud, Tarek and Che, Meiru and Yang, Guowei},
  booktitle={Proceedings of the ACM/IEEE 44th International Conference on Software Engineering: Companion Proceedings},
  pages={1--5},
  year={2022}
}

@inproceedings{yi2020summary,
  title={Summary-guided incremental symbolic execution},
  author={Yi, Qiuping and Wen, Junye and Yang, Guowei},
  booktitle={Proceedings of the ACM/IEEE 42nd International Conference on Software Engineering: Companion Proceedings},
  pages={310--311},
  year={2020}
}

@inproceedings{pham2019cradle,
  title={CRADLE: cross-backend validation to detect and localize bugs in deep learning libraries},
  author={Pham, Hung Viet and Lutellier, Thibaud and Qi, Weizhen and Tan, Lin},
  booktitle={2019 IEEE/ACM 41st International Conference on Software Engineering (ICSE)},
  pages={1027--1038},
  year={2019},
  organization={IEEE}
}

@inproceedings{wang2020deep,
  title={Deep learning library testing via effective model generation},
  author={Wang, Zan and Yan, Ming and Chen, Junjie and Liu, Shuang and Zhang, Dongdi},
  booktitle={Proceedings of the 28th ACM Joint Meeting on European Software Engineering Conference and Symposium on the Foundations of Software Engineering},
  pages={788--799},
  year={2020}
}

@inproceedings{guo2020audee,
  title={Audee: Automated testing for deep learning frameworks},
  author={Guo, Qianyu and Xie, Xiaofei and Li, Yi and Zhang, Xiaoyu and Liu, Yang and Li, Xiaohong and Shen, Chao},
  booktitle={Proceedings of the 35th IEEE/ACM International Conference on Automated Software Engineering},
  pages={486--498},
  year={2020}
}

@inproceedings{liu2023nnsmith,
  title={Nnsmith: Generating diverse and valid test cases for deep learning compilers},
  author={Liu, Jiawei and Lin, Jinkun and Ruffy, Fabian and Tan, Cheng and Li, Jinyang and Panda, Aurojit and Zhang, Lingming},
  booktitle={Proceedings of the 28th ACM International Conference on Architectural Support for Programming Languages and Operating Systems, Volume 2},
  pages={530--543},
  year={2023}
}

@inproceedings{gu2022muffin,
  title={Muffin: Testing deep learning libraries via neural architecture fuzzing},
  author={Gu, Jiazhen and Luo, Xuchuan and Zhou, Yangfan and Wang, Xin},
  booktitle={Proceedings of the 44th International Conference on Software Engineering},
  pages={1418--1430},
  year={2022}
}

@inproceedings{wei2022free,
  title={Free lunch for testing: Fuzzing deep-learning libraries from open source},
  author={Wei, Anjiang and Deng, Yinlin and Yang, Chenyuan and Zhang, Lingming},
  booktitle={Proceedings of the 44th International Conference on Software Engineering},
  pages={995--1007},
  year={2022}
}

@inproceedings{xie2022docter,
  title={DocTer: documentation-guided fuzzing for testing deep learning API functions},
  author={Xie, Danning and Li, Yitong and Kim, Mijung and Pham, Hung Viet and Tan, Lin and Zhang, Xiangyu and Godfrey, Michael W},
  booktitle={Proceedings of the 31st ACM SIGSOFT International Symposium on Software Testing and Analysis},
  pages={176--188},
  year={2022}
}

@inproceedings{deng2022fuzzing,
  title={Fuzzing deep-learning libraries via automated relational api inference},
  author={Deng, Yinlin and Yang, Chenyuan and Wei, Anjiang and Zhang, Lingming},
  booktitle={Proceedings of the 30th ACM Joint European Software Engineering Conference and Symposium on the Foundations of Software Engineering},
  pages={44--56},
  year={2022}
}

@inproceedings{deng2023largetitan,
  title={Large language models are zero-shot fuzzers: Fuzzing deep-learning libraries via large language models},
  author={Deng, Yinlin and Xia, Chunqiu Steven and Peng, Haoran and Yang, Chenyuan and Zhang, Lingming},
  booktitle={Proceedings of the 32nd ACM SIGSOFT international symposium on software testing and analysis},
  pages={423--435},
  year={2023}
}

@article{miotto2018deep,
  title={Deep learning for healthcare: review, opportunities and challenges},
  author={Miotto, Riccardo and Wang, Fei and Wang, Shuang and Jiang, Xiaoqian and Dudley, Joel T},
  journal={Briefings in bioinformatics},
  volume={19},
  number={6},
  pages={1236--1246},
  year={2018},
  publisher={Oxford University Press}
}

@article{heaton2017deep,
  title={Deep learning for finance: deep portfolios},
  author={Heaton, James B and Polson, Nick G and Witte, Jan Hendrik},
  journal={Applied Stochastic Models in Business and Industry},
  volume={33},
  number={1},
  pages={3--12},
  year={2017},
  publisher={Wiley Online Library}
}

@INPROCEEDINGS{xamt,
  author={Duan, Bin and Dong, Ruican and Dong, Naipeng and Kim, Dan Dongseong and Yang, Guowei},
  booktitle={2025 IEEE 36th International Symposium on Software Reliability Engineering (ISSRE)}, 
  title={XAMT: Cross-Framework API Matching for Testing Deep Learning Libraries}, 
  year={2025},
  volume={},
  number={},
  pages={191-202},
  doi={10.1109/ISSRE66568.2025.00030}}

@article{li2018fuzzing,
  title={Fuzzing: a survey},
  author={Li, Jun and Zhao, Bodong and Zhang, Chao},
  journal={Cybersecurity},
  volume={1},
  number={1},
  pages={1--13},
  year={2018},
  publisher={SpringerOpen}
}

@inproceedings{chen2019history,
  title={History-guided configuration diversification for compiler test-program generation},
  author={Chen, Junjie and Wang, Guancheng and Hao, Dan and Xiong, Yingfei and Zhang, Hongyu and Zhang, Lu},
  booktitle={2019 34th IEEE/ACM International Conference on Automated Software Engineering (ASE)},
  pages={305--316},
  year={2019},
  organization={IEEE}
}

@inproceedings{holler2012fuzzing,
  title={Fuzzing with code fragments},
  author={Holler, Christian and Herzig, Kim and Zeller, Andreas},
  booktitle={21st USENIX Security Symposium (USENIX Security 12)},
  pages={445--458},
  year={2012}
}

@article{deng2023large,
  title={Large language models are edge-case fuzzers: Testing deep learning libraries via fuzzgpt},
  author={Deng, Yinlin and Xia, Chunqiu Steven and Yang, Chenyuan and Zhang, Shizhuo Dylan and Yang, Shujing and Zhang, Lingming},
  journal={arXiv preprint arXiv:2304.02014},
  year={2023}
}

@inproceedings{rao2018deep,
  title={Deep learning for self-driving cars: Chances and challenges},
  author={Rao, Qing and Frtunikj, Jelena},
  booktitle={Proceedings of the 1st international workshop on software engineering for AI in autonomous systems},
  pages={35--38},
  year={2018}
}

@inproceedings{bogdoll2022anomaly,
  title={Anomaly detection in autonomous driving: A survey},
  author={Bogdoll, Daniel and Nitsche, Maximilian and Z{\"o}llner, J Marius},
  booktitle={Proceedings of the IEEE/CVF conference on computer vision and pattern recognition},
  pages={4488--4499},
  year={2022}
}

@article{arabahmadi2022deep,
  title={Deep learning for smart Healthcare—A survey on brain tumor detection from medical imaging},
  author={Arabahmadi, Mahsa and Farahbakhsh, Reza and Rezazadeh, Javad},
  journal={Sensors},
  volume={22},
  number={5},
  pages={1960},
  year={2022},
  publisher={MDPI}
}

@article{venkateswarlu2022efficient,
  title={An efficient outlier detection with deep learning-based financial crisis prediction model in big data environment},
  author={Venkateswarlu, Yalla and Baskar, K and Wongchai, Anupong and Gauri Shankar, Venkatesh and Paolo Martel Carranza, Christian and Gonz{\'a}les, Jos{\'e} Luis Arias and Murali Dharan, AR},
  journal={Computational Intelligence and Neuroscience},
  volume={2022},
  year={2022},
  publisher={Hindawi}
}

@inproceedings{christou2023ivysyn,
  title={$\{$IvySyn$\}$: Automated Vulnerability Discovery in Deep Learning Frameworks},
  author={Christou, Neophytos and Jin, Di and Atlidakis, Vaggelis and Ray, Baishakhi and Kemerlis, Vasileios P},
  booktitle={32nd USENIX Security Symposium (USENIX Security 23)},
  pages={2383--2400},
  year={2023}
}

@inproceedings{wang2022eagle,
  title={EAGLE: creating equivalent graphs to test deep learning libraries},
  author={Wang, Jiannan and Lutellier, Thibaud and Qian, Shangshu and Pham, Hung Viet and Tan, Lin},
  booktitle={Proceedings of the 44th International Conference on Software Engineering},
  pages={798--810},
  year={2022}
}

@article{xiao2022metamorphic,
  title={Metamorphic testing of deep learning compilers},
  author={Xiao, Dongwei and Liu, Zhibo and Yuan, Yuanyuan and Pang, Qi and Wang, Shuai},
  journal={Proceedings of the ACM on Measurement and Analysis of Computing Systems},
  volume={6},
  number={1},
  pages={1--28},
  year={2022},
  publisher={ACM New York, NY, USA}
}

\end{document}